# Impact of the NO annealing duration on the SiO$_2$/4H-SiC interface properties in lateral MOSFETs: the energetic profile of the near-interface-oxide traps

Patrick Fiorenza[1*], Marco Zignale[1], Marco Camalleri[2], Laura Scalia[2], Edoardo Zanetti[2], Mario Saggio[2], Filippo Giannazzo[1], Fabrizio Roccaforte[1]

[1]Consiglio Nazionale delle Ricerche – Istituto per la Microelettronica e Microsistemi (CNR-IMM), Strada Ottava 5, Zona Industriale, 95121 Catania, Italy

[2] STMicroelectronics, Stradale Primosole 50, 95121 Catania, Italy

[*]patrick.fiorenza@imm.cnr.it

**Abstract.** In this work, the effects of the duration of the post deposition annealing (PDA) in nitric oxide (NO) on the properties of SiO$_2$/4H-SiC interfaces in n-channel lateral MOSFETs are investigated, with a special focus on the modifications of the energy profile of near-interface-oxide traps (NIOTs). For this purpose, the electrical characteristics of lateral MOSFETs were studied in strong inversion conditions, monitoring the threshold voltage variations due to charge trapping effects. To determine the energetic position of the NIOTs with respect of the SiO$_2$ conduction band edge, the Fermi level position in the insulating layer was evaluated by TCAD simulations of the band diagrams. PDAs of the gate oxide of different duration resulted into similar shape of the energetic profile of the traps inside the insulator with respect of the SiO$_2$ conduction band edge, but with different magnitude. Finally, the effective decrease of the insulator traps is demonstrated despite a saturation of the interface state density under prolonged PDAs and in particular the charge trapped at the NIOTs is reduced from 1-2 $\times$ 10$^{11}$ cm$^{-2}$ down to 3 $\times$ 10$^{11}$ cm$^{-2}$ varying the PDA duration from 10 up to 120 min in NO at 1175 °C.

**Introduction**

Today, 4H-SiC metal oxide semiconductor field effect transistors (4H-SiC MOSFETs) are becoming dominant players in the field of energy conversion systems for electrical vehicles and renewable energies [1,2,3,4]. However, in spite of the rapid progresses made in 4H-SiC power MOSFETs technology, additional efforts are still needed for a deeper comprehension of the physical phenomena associated to the transport in the inversion channel and to further improve the device performances towards the ideal limit.

The conduction behavior of 4H-SiC MOSFETs is known to be strongly influenced by the processing of the $SiO_2$/4H-SiC interface [5,6]. In particular, the on-resistance and the field effect channel mobility ($\mu_{FE}$) [7,8] are typically optimized by employing post-oxidation-annealing (POA) or post-deposition-annealing (PDA) of the $SiO_2$ gate insulator in nitric oxide (NO) [9]. However, if on one side the introduction of nitrogen during PDA improves the channel mobility through the passivation of interface states, on another side it can also generate trapping states at the $SiO_2$/4H-SiC interface, which in turn may have a detrimental impact on the stability of the threshold voltage ($V_{th}$) [10,11]. In addition, the presence of these trapping states, inside the insulator and in the proximity of the $SiO_2$/4H-SiC interface, also affects the on-state conduction in n-channel MOSFETs. In fact, free electrons in the channel can be trapped either in the deep interface states, e.g. close to the 4H-SiC valence band edge, or inside the insulator in the near interface region (NIOTs) [12]. In this context, it is mandatory to explore the evolution, under different PDA conditions, of the interface states density ($D_{it}$) not only close to 4H-SiC conduction band, but also inside the insulating layer in the proximity of the channel region. Moreover, it is also relevant to monitor the energy distribution profile of the NIOTs that might be generated and/or modified under prolonged PDAs in NO. *Hatakeyama et al.* [13] recently reported on the Hall mobility degradation in 4H-SiC MOSFETs subjected to prolonged PDAs in NO. However, the evolution of the NIOTs as a function of the PDA duration was not discussed in that paper.

Following the initial work of *Afanas'ev et al.* [14], in recent years significant efforts have been focused on the study of the nature of the NIOTs [15,16], both below and above the 4H-SiC conduction band edge [17]. In addition, several approaches for the determination of the energy distribution of NIOTs have been reported. In particular, *Hauck et al.* [15] presented a Hall-effect based method to empirically determine the NIOTs energetic profiles in lateral MOSFETs fabricated with POA in $O_2$ and NO, respectively. On the other hand, *Pande et al.* [16] evaluated the energetic profile of active NIOTs using AC-current measurements on MOS capacitors under strong accumulation. However, up to now a methodology to determine the NIOTs energetic profile in 4H-SiC MOSFETs was not uniquely defined.

Based on all these literature findings, it is clear that the energetic distribution of the NIOTs inside the insulating layer for different PDAs duration in NO deserves to be investigated. In this way, it can be established if new NIOTs are introduced during prolonged PDA in NO, which can be detrimental for the device performances.

In this paper, we firstly applied a cyclic gate bias stress procedure to separate the contributions of the trapping states at the interface (close to the band edges) from the NIOTs. Therefore the energetic position of the NIOTs with respect to the $SiO_2$ conduction band edge was determined using the TCAD simulated band diagrams of the channel region at different gate biases. Finally, the NIOTs energetic profiles for PDAs with different duration were compared. The results were discussed by the comparison of the interface state density profile evolution getting insights on the NIOTs trapping mechanism and energetic profile modification under PDAs in NO.

**Experimental**

In our experiments, test-devices consisting in n-channel lateral MOSFETs fabricated on 4H-SiC were used. In these devices, a n-type epitaxial layer with a donor concentration $N_D = 9 \times 10^{15}$ cm$^{-3}$ was implanted with aluminum (Al) ions at different energies and fluencies in order to have an almost flat profile, and subjected to a high-temperature activation annealing [18]. The resulting acceptor density of the body region was $N_A \sim 2\text{-}3 \times 10^{17}$ cm$^{-3}$. The $SiO_2$ gate oxide was deposited by a low-pressure chemical vapor deposition (LPCVD) furnace using dichlorosilane (DCS) [19] and had a thickness of 55 nm. Then, PDAs of different duration (from 10 to 120 minutes) were carried out at 1175 °C in NO in a horizontal furnace in a sub-atmospheric pressure regime [19].

Fig.1 schematically shows the processing temperature ramp used for the PDA of the MOSFETs. In particular, after a heating ramp up to 1175 °C in $N_2$, a stabilization step of 20 min is performed at this temperature (pre-PDA). Then, the PDA is performed with different time duration, from 10 to 120 minutes. Finally, a purge step of 10 min in $N_2$ is performed before the ramp down to the extraction temperature (900 °C).

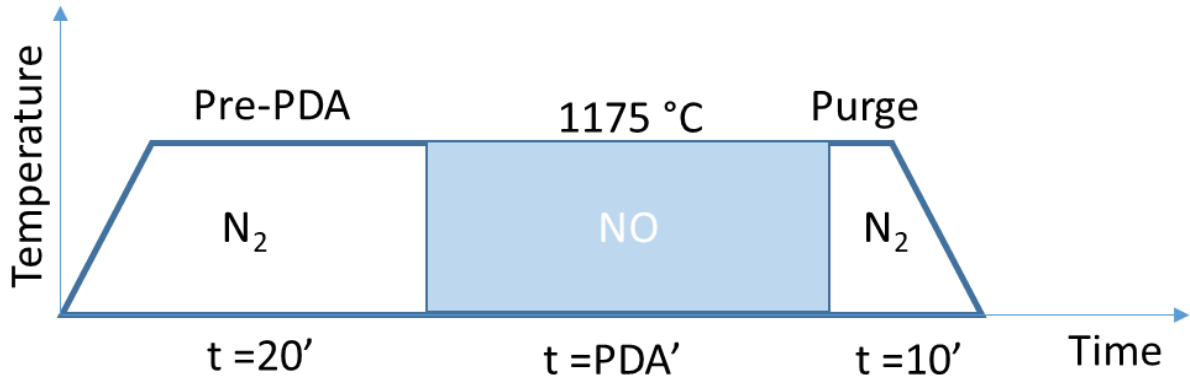

Fig.1: Schematic representation of the process temperature ramp of the PDA carried on the lateral 4H-SiC MOSFETs. The PDA duration in NO varied between 10 and 120 minutes.

The electrical characterization of the fabricated 4H-SiC MOSFETs was carried out in a CASCADE Microtech probe station equipped with a Keysight B1505A parameter analyzer, acquiring the current voltage ($I_D$-$V_G$) transfer characteristics and capacitance–voltage (C-V) curves.

The TCAD simulations of energy band-diagrams in the channel region were carried out with the process tools Sentaurus by Synopsys [20] using 4H-SiC and $SiO_2$ ideal parameters library.

**Results and discussion**

Firstly, the inversion channel conduction properties – as an indication of the MOSFETs $R_{ON}$ conduction properties – was obtained considering the device maximum value transconductance in the linear region by equation: [21]

$$g_m = \frac{\partial I_D}{\partial V_{GS}} \tag{1}$$

with a fixed $V_{DS}$ source-drain potential the derivative $\partial I_D/\partial V_{GS}$ is the MOSFET transconductance.

In fact, the $g_m$ value gives qualitative information on the modulation of the channel conductivity by the application of the gate bias being also linked by text book equation to the field effect mobility $\mu_{FE}$.

Fig. 2 reports the collection of the field effect mobility $g_m$ values as a function of the gate bias $V_G$, measured on the 4H-SiC MOSFETs subjected to different PDAs in NO for 10, 20, 50, 90, and 120 min. As can be noticed, first the $\mu_{FE}$ values increased with increasing the PDA duration, but after 50 min of PDA their values is increased of few units, reaching a value of about $4\times10^{-7}$ S at the gate bias $V_G = 20V$. From these values it is possible to derive the $\mu_{FE}$ maximum values that are reported in Table I.

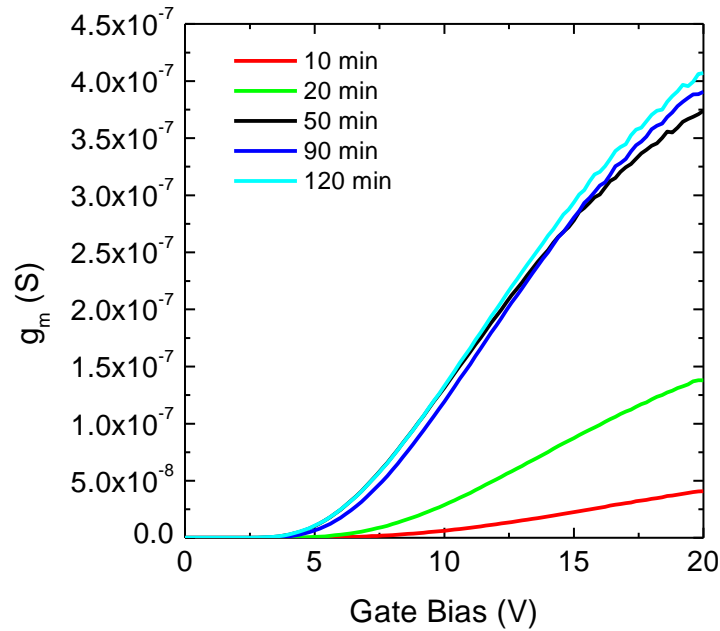

Fig.2: Devices transconductance $g_m$ as a function of the gate bias $V_G$ determined in 4H-SiC lateral MOSFETs subjected to PDAs in NO at 1175°C of different duration from 10 to 120 min.

Fig.3a shows the C-V curves collected at 1 kHz with a sweep rate of 0.1V/s from negative toward positive gate bias values and backwards. Under these measurement conditions, a small hysteresis is visible in the C-V curves due the charging of deep hole traps in the 4H-SiC band gap [22,23,24]. An average insulator thickness value of about 55 nm was estimated from the capacitance values in accumulation condition $C_{ox}$. Furthermore, from the $C_{ox}$ variation for different PDA durations, the oxide thickness variation can be estimated to be less than 2 nm for PDA times longer than 20 min. Hence, it can be concluded that only a limited substrate re-oxidation occurs during the different NO annealing durations chosen in our experiments.

However, as can be noticed in Fig.3a, the slope of the C-V curves – reaching the inversion condition for $V_G > 0V$ – gradually increases with increasing the NO annealing duration, approaching the slope of the ideal curve for the lateral MOSFET simulated by TCAD. The C-V curves collected on the lateral MOSFETs were used to extrapolate the $D_{it}$ profile close to the 4H-SiC conduction band edge. It should be noticed that, differently from the simple MOS capacitors typically used for $D_{it}$ extraction, the lateral MOSFET exposes at the $SiO_2$/4H-SiC interface simultaneously p-type and n-type materials, where the surface potential has different values. Hence the TCAD simulated ideal C-V curve for the lateral MOSFET was used as a reference to extract the $D_{it}$ profile by the Terman's method [25], where the low frequency (1 kHz) experimental curve is compared with the ideal one. As shown in Fig. 3b, the $D_{it}$ initially decreases with increasing the NO annealing duration up to 50 min, and therefore saturates, consistently with the previously reported behavior of the mobility (see Fig.2).

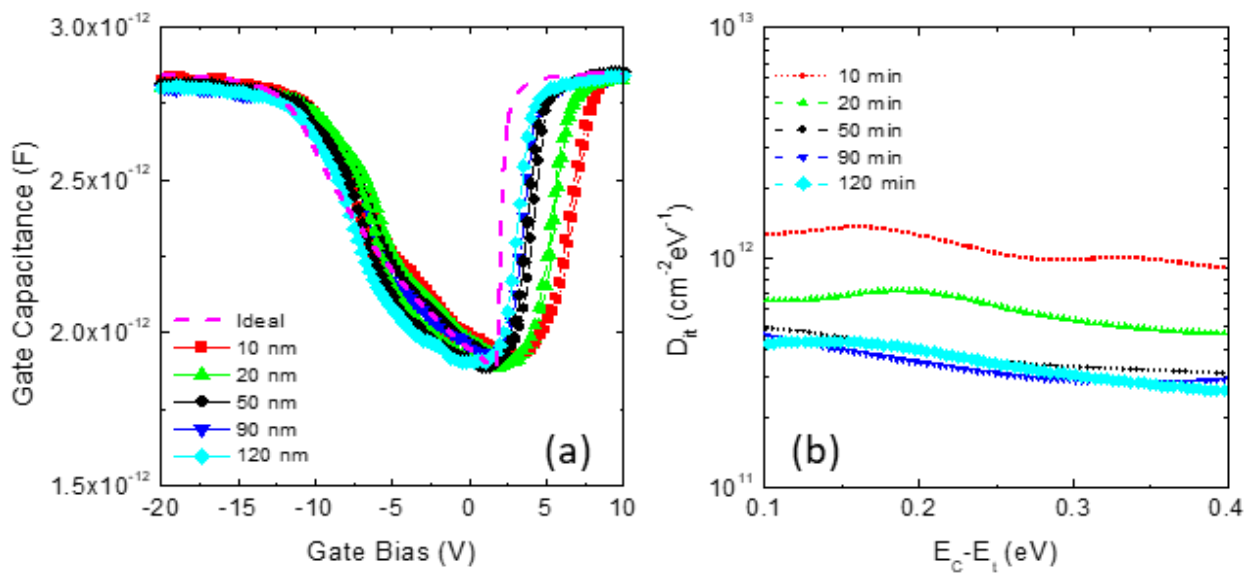

Fig.3: (a) C-V curves measured at 1 kHz on 4H-SiC MOSFETs subjected to PDAs in NO at 1175°C of different duration compared to the ideal curve (dashed line). (b) $D_{it}$ values for the same devices extracted using the Terman's method.

The effect of the NO annealing duration on the threshold voltage $V_{th}$ stability was also studied. In fact, it has been previously observed that a prolonged exposure to oxidizing species (such as the NO) may increase the interfacial disorder, due to the formation of sub-stoichiometric oxide layer and/or carbon-related defects [19]. In our case, by applying the non-relaxing stressing method to study the MOSFET $V_{th}$ stability described in Ref. [9], it was possible to separate the different trapping contributions characterizing the $SiO_2$/4H-SiC system under investigation. The method is based on a cyclic ramped gate bias stress procedure followed by a single point drain current measurement. This procedure is used to probe the interface and the near interface traps in the $SiO_2$/4H-SiC system over the whole 4H-SiC bandgap. In particular, by applying a positive gate bias stress ($V_G > +5V$), the MOSFET channel is inverted and a large amount of electrons is attracted toward the $SiO_2$/4H-SiC interface due to the crossing of the Fermi level above the semiconductor conduction band. Hence, under these bias conditions, it is possible to stimulate the NIOTs that are energetically located above the 4H-SiC conduction band edge in a narrow region of the insulating layer close to the $SiO_2$/4H-SiC interface. The threshold voltage variation $\Delta V_{th}$ observed after this gate bias stress was evaluated by a single point measure of the root square of the drain current ($I_D^{1/2}$), assuming the linear approximation described in our previous work [9].

The threshold voltage variation $\Delta V_{th}$ as a function of the gate bias stress in inversion regime, i.e. when the Fermi level is above the semiconductor conduction band, is reported in Fig.4 for the different annealing conditions. The arrows in Fig. 4 indicate the gate bias stress direction. The gate bias stress is applied for 2 seconds at each measurement point and it started from 0 V up to + 30 V with a step of + 5 V keeping the MOSFET channel in the inversion condition. Afterwards, the gate bias stress is diminished from + 30 V down to 0 V with a step of + 5 V. The single drain current measurements is converted in the $\Delta V_{th}$ values, and then the $\Delta V_{th}$ values are converted into the amount of trapped charge ($N_{Trap}$) in the NIOTs above the 4H-SiC conduction ($E_C$) using the following relation:

$$N_{Trap} = \frac{\kappa \varepsilon_0 \Delta V_{th}}{q t_{ox}} \qquad (2)$$

where q is the electron charge, $t_{ox}$ and $\kappa$ are the SiO$_2$ thickness and relative permittivity, respectively, and $\varepsilon_0$ is the vacuum permittivity.

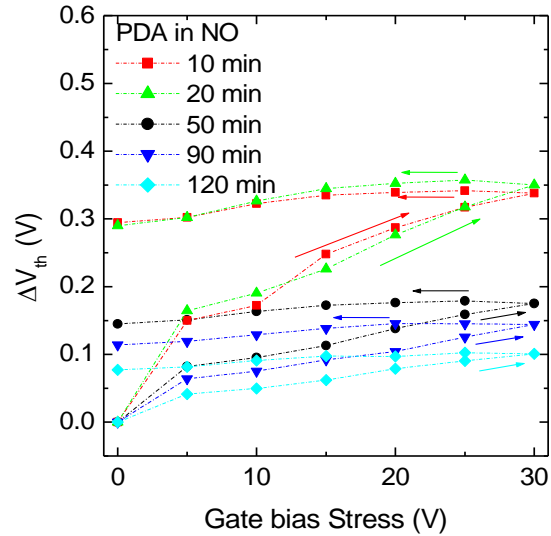

Fig.4: Threshold voltage variation $\Delta V_{th}$ measured as a function of the gate bias stress on the 4H-SiC MOSFETs subjected to PDAs in NO of different duration. The arrows indicate the direction of the stress.

Table I summarizes the electrical results obtained from the I-V can C-V measurements carried out on the 4H-SiC MOSFETs, reporting the values of field effect mobility $\mu_{FE}$ acquired at $V_G=20V$, the interface states density $D_{it}$ taken at 0.1 eV below the 4H-SiC conduction band, and the amount of trapped charge ($N_{Trap}$) in the NIOTs.

Table I. Summary of the electrical data collected extracted on the investigated 4H-SiC MOSFETs for different PDAs duration in NO at 1175°C.

| Sample (PDA duration) | $\mu_{FE}$ (cm$^2$V$^{-1}$s$^{-1}$) | $D_{it}$ (×10$^{11}$ cm$^{-2}$eV$^{-1}$) | $N_{Trap}$ (×10$^{11}$ cm$^{-2}$) |
|---|---|---|---|
| 1 (10 min) | 4.5 | 12.48 | 1.23 |
| 2 (20 min) | 12.4 | 6.51 | 1.26 |
| 3 (50 min) | 33.8 | 4.21 | 0.65 |
| 4 (90 min) | 34.5 | 4.56 | 0.53 |
| 5 (120 min) | 35.0 | 4.21 | 0.37 |

As can be noticed, the increase of the PDA duration results in an increase of the field effect mobility (Fig. 2) and a reduction of the $D_{it}$ (Fig. 3), but both values saturated the improvement after 50 min of PDA in NO. On the other hand, the amount of charge trapped in the NIOTs ($N_{Trap}$) decreases with increasing the PDA duration (derived from Fig. 4 and Eq. 2), with no saturation even after 120 min of PDA in NO.

In order to get further insights on the physics of the NIOTs above the 4H-SiC conduction band edge, numerical TCAD simulation were employed to evaluate the MOSFET conduction band diagram for different positive gate bias values, as illustrated in Fig. 5. The simulation is carried out taking into account the Synopsys library Fermi statistics, Shockley-Read-Hall recombination, and the following physical parameters: $SiO_2$ and 4H-SiC electron affinities $\chi_{SiO_2} = 0.9$ eV and $\chi_{4H-SiC} = 3.6$ eV; metal gate work function of 4.15 eV; $SiO_2$ and 4H-SiC band gaps (at T = 0 K) of $E_{g,SiO_2} = 9$ eV and $E_{g,4H-SiC}$ 3.285 eV, respectively. The simulation results allowed us to evaluate the $SiO_2$/4H-SiC conduction band discontinuity for each positive gate voltage value applied in the cyclic bias stress. Fig. 5 illustrates a particular region that extends from the $SiO_2$/4H-SiC interface 5 nm inside the insulator side and 10 nm inside the semiconductor substrate, respectively. The energy axis is referred to the position of the Fermi level, which is conventionally placed at zero, independently of the gate bias value. As can be noticed, the Fermi level lies always above the 4H-SiC conduction band edge, thus meaning the formation of an inverted electron layer at the $SiO_2$/4H-SiC interface. Once the inversion layer is formed, it can be argued that electrons move throughout the insulator layer according to the tunneling probability. However, if some traps are present inside the insulator near the interface region, electrons can be trapped in these NIOTs. By increasing the positive applied gate bias, the $SiO_2$ downwards band bending becomes more pronounced, thus enabling the energetic alignment between the trapping states in the upper part of the $SiO_2$ band gap with the Fermi level. Hence, by increasing the gate bias, it is possible to reduce the energetic distance between the $SiO_2$ conduction band edge and the Fermi level. This information is used to determine the amount of charge trapped in that specific position converting the $\Delta V_{th}$ into $N_{Trap}$ as described before.

Fig. 6 shows the calculated conduction band position with respect of the Fermi level $E_F$ across the SiO$_2$/4H-SiC interface for different gate bias value. $E_F$ is kept at 0 eV. As can be seen, for positive gate bias, the $E_C^{4H\text{-}SiC} - E_F$ value becomes negative, thus meaning that the Fermi level lies above the 4H-SiC conduction band edge. The position of the Fermi level with the respect of the 4H-SiC conduction band edge can be used to determine the corresponding energetically aligned traps inside the SiO$_2$ layer. In fact, the offset between the semiconductor and insulator conduction band edges is estimated considering the ideal $\Delta E_C$ value commonly used in literature [26,27] of 2.7 eV. Hence, data reported in Fig. 6 can be used to extract the energetic position of the $N_{Trap}$ referred to the SiO$_2$ conduction band edge. In fact, keeping the Fermi level at $E = 0$ eV as reference, it is possible to evaluate which are the NIOTs traps aligned with the Fermi level in the semiconductor. Such aligned NIOTs contribute to the $N_{Trap}$, and their energy position from the SiO$_2$ conduction band edge can be calculated according to the relation:

$$E_C^{SiO_2} - E_F = \Delta E_C - (E_C^{4H\text{-}SiC} - E_F). \tag{3}$$

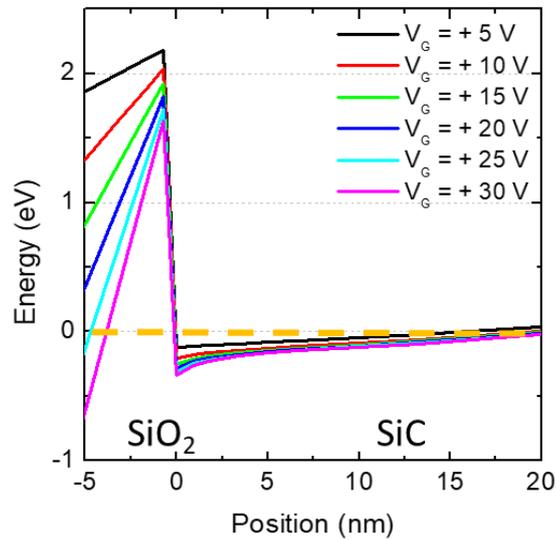

Fig.5: Calculated conduction and valence band of 4H-SiC at different gate bias values. The Fermi level across the SiO$_2$/4H-SiC interface is located at 0 eV.

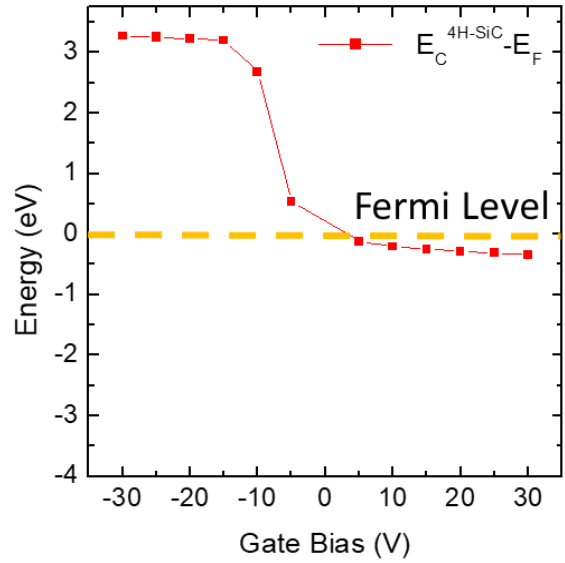

Fig.6: Calculated 4H-SiC conduction band position with respect of the Fermi level $E_F$ across the $SiO_2$/4H-SiC interface for different gate bias values. The Fermi level $E_F$ across the $SiO_2$/4H-SiC interface is put at 0 eV.

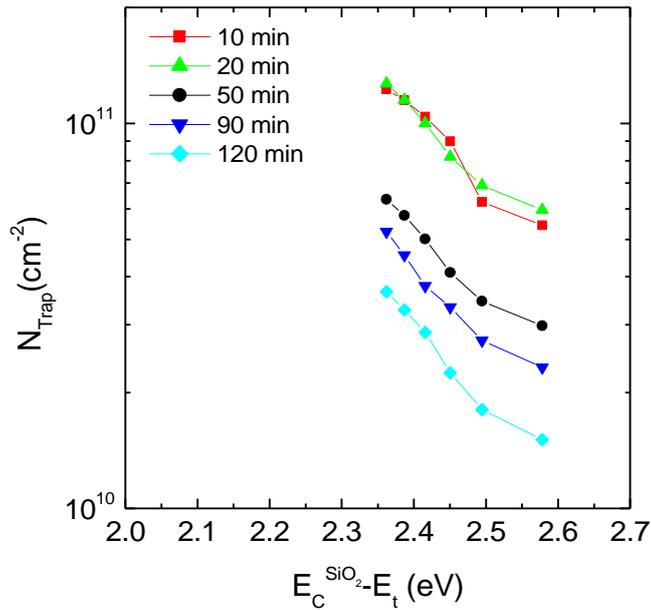

Fig.7: NIOTs energetic profiles for the 4H-SiC MOSFETs subjected to PDAs in NO at 1175°C of different duration.

Fig. 7 shows the NIOTs energetic profiles, referred to the $SiO_2$ conduction band edge, estimated by the simultaneous use of the $\Delta V_{th}$ values obtained by the gate bias stress (Fig.4) and the calculated conduction band edge with respect of the Fermi level (Fig. 6) using the equation (3).

As can be seen, PDAs for 10 and 20 min result into larger amount of trapped charges in the NIOTs compared to the other annealing conditions. On the other hand, the charges trapped in the NIOTs decreases with increasing the PDAs duration. However, in all cases the NIOTs energetic profiles exhibit the same decreasing trend with increasing the energy depth from the SiO$_2$ conduction band edge.

As proposed by *A. Lelis et al.* [28], if the charging mechanism of the NIOTs is assisted by the presence of interface states, it is reasonable to suppose that the detected N$_{Trap}$ at the NIOTs are influenced by the D$_{it}$ profile. However, in our case the D$_{it}$ are nearly constant in the samples subjected to the longer PDAs for 50, 90 and 120 min (Fig. 3b). Hence, the total decreasing amount of detected N$_{Trap}$ at the NIOTs measured by increasing the PDAs duration can be related to a decrease of the defects in the insulator.

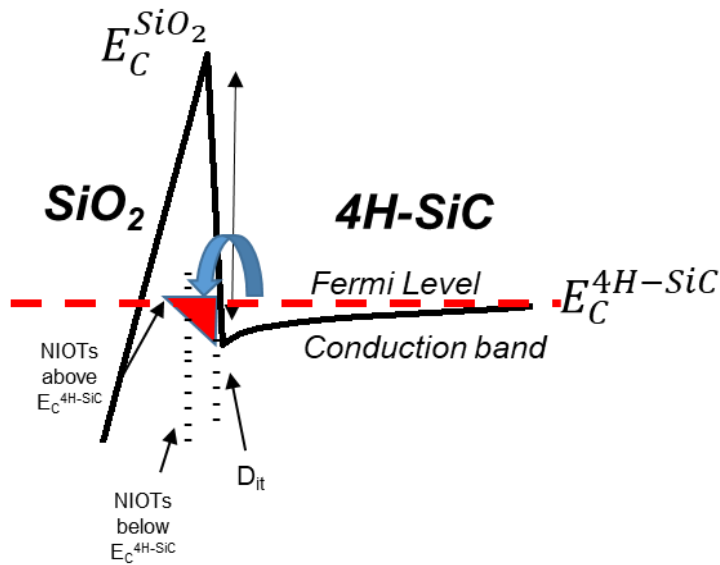

Fig.8: Graphical representation of the SiO$_2$/4H-SiC interface in 4H-SiC MOSFET under positive gate bias, indicating the conduction band discontinuity and the presence of interface states and oxide traps both above and below the 4H-SiC conduction band edge.

Fig. 8 illustrates graphically the situation of the SiO$_2$/4H-SiC interface in 4H-SiC MOSFET under positive gate bias, reporting the energetic position of the NIOTs both above and below the E$_C^{4H-SiC}$ edge. In particular, the ΔE$_C$ off-set between the SiO$_2$ and 4H-SiC conduction band edges is used to energetically locate the NIOTs above and below the E$_C^{4H-SiC}$ edge. Furthermore, it can be concluded that NIOTs below the E$_C^{4H-SiC}$ edge affect the inversion channel conductivity either with Coulomb scattering and/or reducing the amount of free carriers. On the other hand, NIOTs above the $E_C^{4H-SiC}$ edge affect the $V_{th}$ stability.

Fig. 9 shows the summary of the results obtained in terms of field effect mobility $\mu_{FE}$, the interface state density $D_{it}$ (at 0.2 eV below the SiC conduction band) and the maximum of trapped charge in the NIOTs. As can be noticed, by increasing the PDA duration, the decrease of the interface states $D_{it}$ is accompanied by an increase of the field effect mobility $\mu_{FE}$. For PDAs longer than 50 min both electrical parameters are saturated. On the other hand, the amount of trapped charge in the NIOTs is decreasing also for long PDAs up to 120 min. Hence a prolonged PDA improved the trapping states inside the insulator.

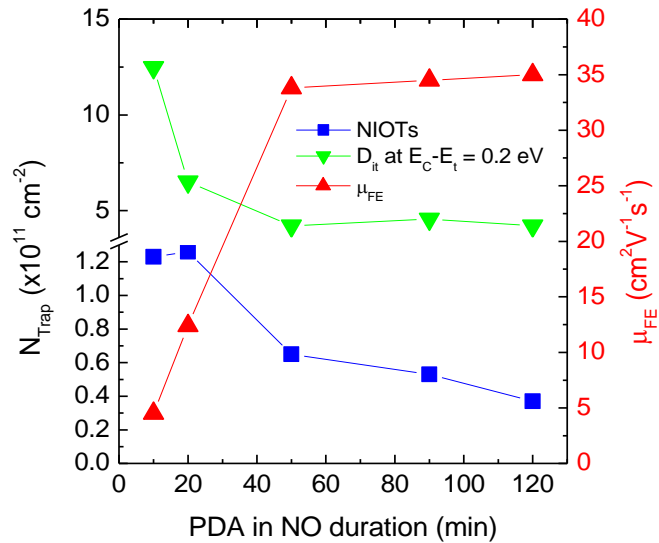

Fig.9: Summary of the electrical parameters obtained by the characterization of the lateral MOSFETs where the SiO$_2$/4H-SiC system was subjected to PDA with different duration. Interface state density, field effect mobility and trapped charge in the NIOTs.

**Conclusion**

In this paper, a combination of numerical simulation and the use of a cyclic high gate bias stress procedure is used to extract the energetic profile of near interface oxide traps (NIOTs) in the proximity of the n-channel of the 4H-SiC lateral MOSFETs subjected to PDAs in NO of different duration. It has been observed that a prolonged PDA in NO resulted in a decrease of the NIOTs while the interface state density reduction is saturated already after short PDA process. These findings were used to determine the NIOTs energetic profile with respect to the $SiO_2$ conduction band edge and its shape is kept similar for all the samples but with a rigid shift in the semi-log scale. These results were obtained by the calculated band diagram of the MOSFET under positive bias was used to estimate the positon of the Fermi level with the respect of the $SiO_2$ conduction band edge. This allowed the definition of the energy position of the electrons trapped in the NIOTs in the samples subjected to different PDAs. In particular, it has been found that the charge trapped at the NIOTs is reduced from $1\text{-}2 \times 10^{11}$ cm$^{-2}$ down to $3 \times 10^{11}$ cm$^{-2}$ varying the PDA duration from 10 up to 120 min in NO at 1175 °C.

The fine control of the NO PDA is required to improve the MOSFET interfacial transport and mitigating the threshold voltage instability.

### Acknowledgements

This paper has been partially supported by Horizon EU Advances in Cost-Effective HV SiC PowerDevices for Europe's Medium Voltage Grids (AdvanSiC). The AdvanSiC project has received funding from the European Union's Horizon Europe programme under grant agreement No 101075709. Views and opinions expressed are however those of the author(s) only and do not necessarily reflect those of the European Union. Neither the European Union nor the granting authority can be held responsible for them.